\documentclass[fleqn,twoside]{article}
\usepackage{espcrc2}

\usepackage{graphicx}
\usepackage[figuresright]{rotating}

\newcommand{\AmS}{{\protect\the\textfont2
  A\kern-.1667em\lower.5ex\hbox{M}\kern-.125emS}}
\def\NPB#1#2#3{{\it Nucl.\ Phys.}\/ {\bf B#1} (#2) #3}

\def\PRL#1#2#3{{\it Phys.\ Rev.\ Lett.}\/ {\bf #1} (#2) #3}

\def\AEF{A.E. Faraggi}

\title{Searching for Extra Dimensions in High Energy Cosmic Rays}

\author{Alessandro Cafarella\address{Dipartimento di Fisica, Universit\`{a} di Lecce and INFN Sezione di Lecce,\\
        Via per Arnesano, 73100 Lecce (Italy)},
        Claudio Corian\`{o}\addressmark\thanks{Presented by C. Corian\`o at the XIII Intl. Symp. on Very High Energy Cosmic Rays Interactions, Pylos, Greece 6-12 Sept. 2004}
        and
        T. N. Tomaras\address{Department of Physics and Institute for Plasma Physics, University of Crete and FORTH,\\
        Heraklion, Crete, Greece}}

\begin{document}

\begin{abstract}
We present a study of the decay of an intermediate mini black hole at the
first impact of a cosmic ray particle with the atmosphere, in the context of 
D-brane world scenarios with TeV scale gravity and large extra dimensions.
We model the decay of the black hole using the semiclassical approximation and
include the corrections coming
from energy loss into the bulk. Extensive simulations
show that mini black hole events are characterized by essentially different multiplicities
and wider lateral distributions of the air showers as a function of the energy of the
incoming primary, as compared to standard events. Implications for their detection and
some open issues on their possible discovery are also briefly addressed.
\vspace{1pc}
\end{abstract}

\maketitle

\section{INTRODUCTION}

At a cursory look, the inclusive cosmic ray spectrum, which has been measured
over a wide range of energy, seems to be pretty simple in its functional shape, characterized 
by a (fast-falling) power-like behaviour over a huge energy range. Several mysteries may be hiding
behind this apparent simplicity.
Two of these mysteries, which have been puzzling theorists and experimentalists 
alike for several decades, and have been extensively discussed in this
Symposium, are the knee
($10^{15}$ eV) and the ankle ($10^{19}$ eV). A long debate which
has spurred from these observations has not reached any final
conclusions as to the origin of the peculiar spectral dependence measured in these regions.
Physics at the knee may involve, according to some interesting
proposals, a critical behaviour of the QCD pomeron \cite{White}, or, 
alternatively, disoriented chiral condensates and production of strangelets or
other exotics, while the existence or the absence of the GZK cutoff \cite{gzk} 
still needs to be completely assessed.
Evading the cutoff using string relics \cite{ccf},\cite{cfp} has 
also been proposed as an interesting possibility.

On the other hand, recent formal suggestions from string theory and 
gauge theory \cite{ED} lead to the exciting possibility of a brane picture of our world,
with a gravity scale lowered from $10^{19}$ GeV down maybe to a fraction of 
a TeV and with the possibility of
envisioning exotic new scenarios where ``black hole resonances'' 
might form even at colliders
such as the LHC \cite{DL}. In cosmic ray physics, in particular,
the energy available at the impact of the primaries with the
atmosphere can be very large and up to several hundreds of TeVs, 
and this points favourably towards a
possible test of these new scenarios \cite{Theodore}. 

We have tested the implications of this hypothesis for cosmic ray 
air showers, by studying, in particular,  the lateral distributions
and the multiplicities of events mediated by an intermediate black hole.
Here we provide a brief summary of our results, while a detailed 
analysis can be found in \cite{CCT}.

\section{MINI BLACK HOLES IN COSMIC RAYS}

Various studies of mini black hole production at the LHC have been 
pulished in the last few years (see \cite{CCT} for a rather 
complete list of references).
At the same time analytical and numerical studies of greybody 
factors and the formation of trapped surfaces in extra dimensional models
have been able to provide grounds for the experimental searches: 
in the context of the brane-world scenario black holes will form 
copiously at hadron colliders, in events characterized by a large 
multiparticle multiplicity and a ``fireball'' structure, features which 
would make this black hole intermediate state look quite distinct from 
other ordinary resonances.

Mini black holes form, according to Thorne's hoop conjecture 
(see \cite{CCT} and references therein),
when the energy in the center of mass of a collision  is concentrated 
into a region of radius 
smaller than the corresponding Schwarzschild radius.
The process may be accompanied by a sizeable amount of 
gravitational energy loss,
with a collision which can be characterized by the scattering of
two gravitational shock waves of Aichelburg-Sexl type at a small 
impact parameter. The so-formed black hole decays
democratically mostly on the brane (our world) into all the 
fundamental states of the standard model model or of the supersymmetric 
standard model, if supersymmetry is also open at that energy scale.

The analysis that we have performed is not based on a Monte Carlo 
modeling of the decay of the black hole \cite{Cavaglia},
rather we write down and compute the istantaneous parton and lepton 
emissions of the decay of the black hole using a
multinomial probability distribution for the hadronization of a 
given parton, and using suitable fragmentation functions. 
These are defined via a renormalization group evolution
up to the mean energy of the immediate decay products of the black
hole evaporation, which is given by the ratio $M_{BH}/N$ of the black 
hole mass $M_{BH}$ divided by the average number $N$ of particles produced
in such a decay, which in turn is determined in the semiclassical
approximation by the entropy of the black hole.

\section{A MODEL FOR A MINI BLACK HOLE DECAY}

As stated above, mini black holes will be produced in the collision 
of any primary cosmic ray particle of appropriate energy with a quark or
gluon of a nucleon in the atmosphere. 
Such a collision may lead to the formation of trapped
surfaces and of an event horizon, whose size can be of the order 
of $10^{-3}$ fm for a fundamental gravity scale $M_*$ of the order of a TeV.
A spherical s-wave forms due to the fragmentation of the black hole and 
the number of states emitted during the decay
is computed according to a standard formula for the multiplicities \cite{DL}, 
which depends on the number of extra dimensions $n$. We work in $D=4+n$ 
spacetime dimensions.

We compute the properties of the decay products, after hadronization, 
by simulating a sequence of uncorrelated air showers using CORSIKA/SIBYLL
\cite{CORSIKA,SIBYLL} and compare the results of the multiplicities and 
the lateral distributions of the various components with benchmark 
simulations of standard events. Other hadronization models are 
also available in CORSIKA, for instance QGSJET \cite{QGSJET}.

\section{RESULTS}

\begin{figure}
{\centering \resizebox*{7cm}{!}{\rotatebox{-90}{\includegraphics{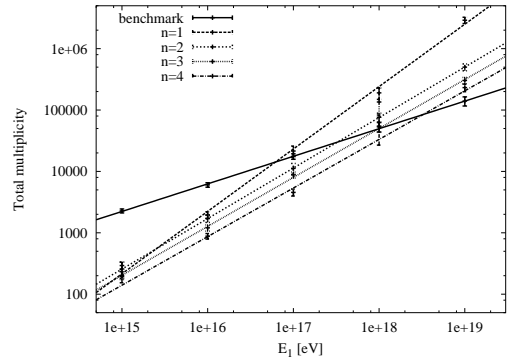}}} \par}

\caption{Plot of the total particle multiplicity
as a function of \protect\( E_{1}\protect \)
The first interaction is kept fixed at \protect\( 5500\, \textrm{m}\protect \)
(517 $g/cm^2$)
and the observation level is at \protect\( 5000\, \textrm{m}\protect \)
(553 $g/cm^2$).
We show in the same plot the benchmark result (with a proton as a primary)
and the mini black hole result for a varying number of extra-dimensions \protect\( n\protect \).\label{fig:bhwl_NrappVE}}
\label{prima}
\end{figure}

\begin{figure}
{\centering \resizebox*{7cm}{!}{\rotatebox{-90}{\includegraphics{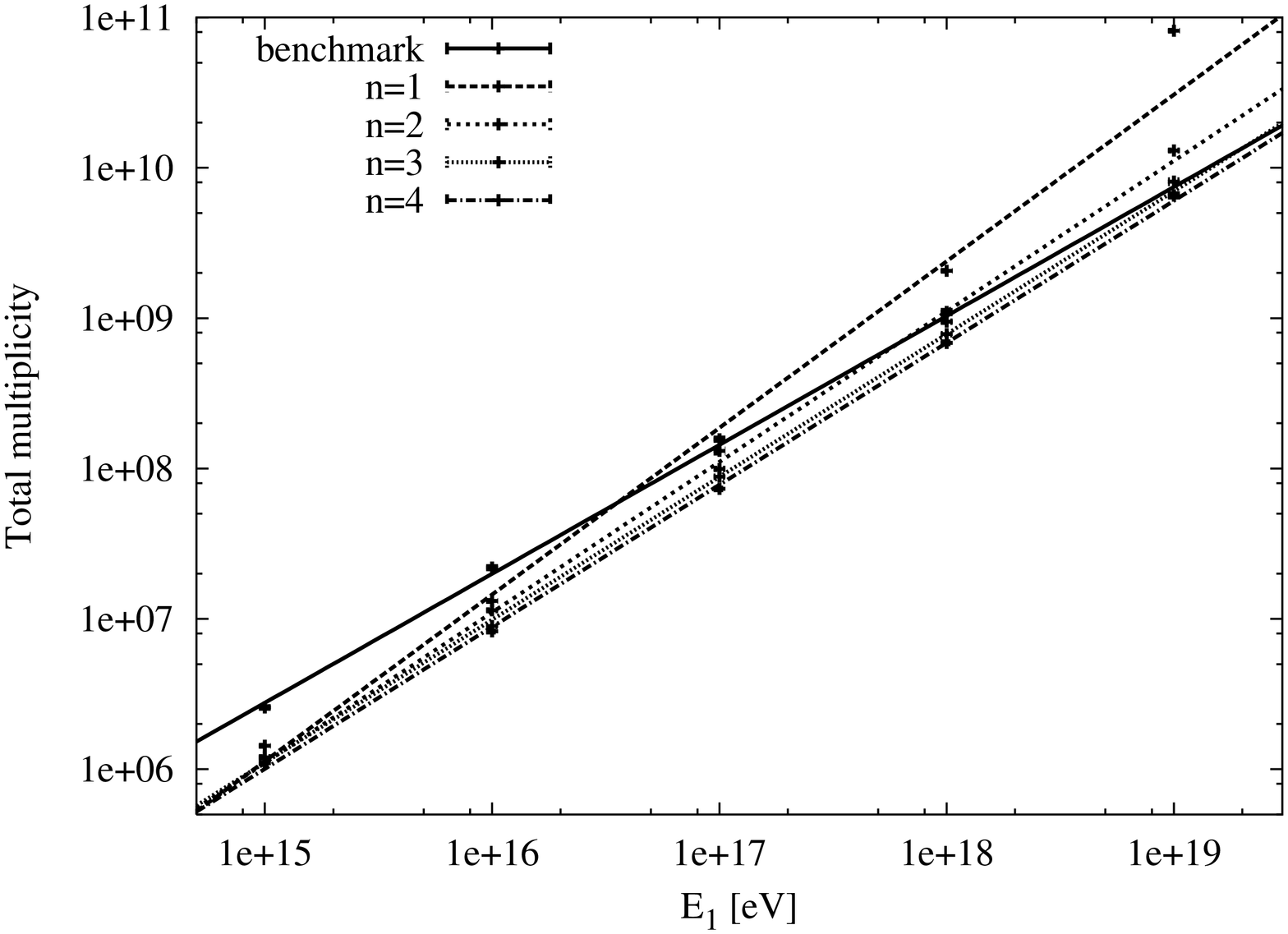}}} \par}
\caption{As in Fig. \ref{prima}, but this time the first interaction occurs at \protect\( 15000\, \textrm{m}\protect \)
(124 $g/cm^2$).}
\label{seconda}
\end{figure}

Simulations of observables of the type discussed in \cite{CCF} 
have been performed at a small cluster. We have summarized in 4 figures 
the most salient results of our studies, where we try to distinguish 
between standard (benchmark) events and black hole mediated events. 
We plot in Fig.~\ref{prima} the total particle multiplicity measured 
at detector level (5000 m) assuming an impact at 5500 m 
(with a depth $X=517$ $g/cm^2$); a similar simulation at a higher altitude 
of 15000 m (with a depth $X=124$ $g/cm^2$) is shown in Fig.~\ref{seconda}. 
At energies of the order of 1000 TeV, relevant to the Centauro events, the 
total multiplicity for $n=4$ is around 100, for a black hole which evaporated 
500 meters above the detector. A black hole produced closer to the detector
will obviously lead to fewer particles in the shower, an overall picture consistent
with the Centauro observations. 
Notice also that the multiplicites of black hole events are lower than in the 
standard events roughly by a
factor of 10 in the "low" energy region, while they are larger also be a factor
of 10 for energies around the GZK cutoff. These differences tend to become
less pronounced for higher altitudes of the initial impact.

\begin{figure}
{\centering \resizebox*{7cm}{!}{\rotatebox{-90}{\includegraphics{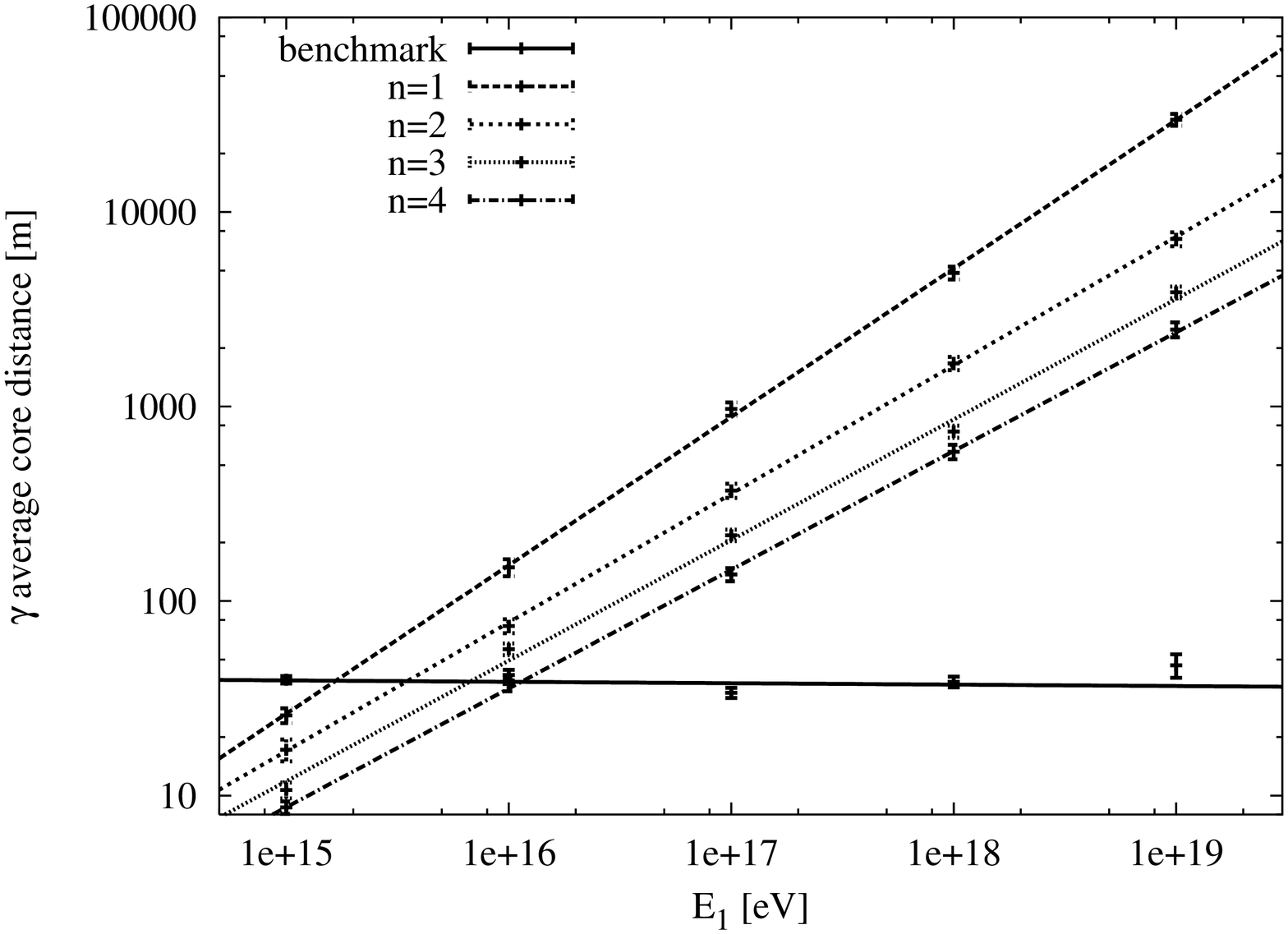}}} \par}
\caption{Plot of the the average
distance $R$ of the core of the shower of photons as a function of \protect\( E_{1}\protect \). The first interaction is kept fixed at \protect\( 5500\, \textrm{m}\protect \)
(517 $g/cm^2$).}
\label{terza}
\end{figure}

\begin{figure}
{\centering \resizebox*{7cm}{!}{\rotatebox{-90}{\includegraphics{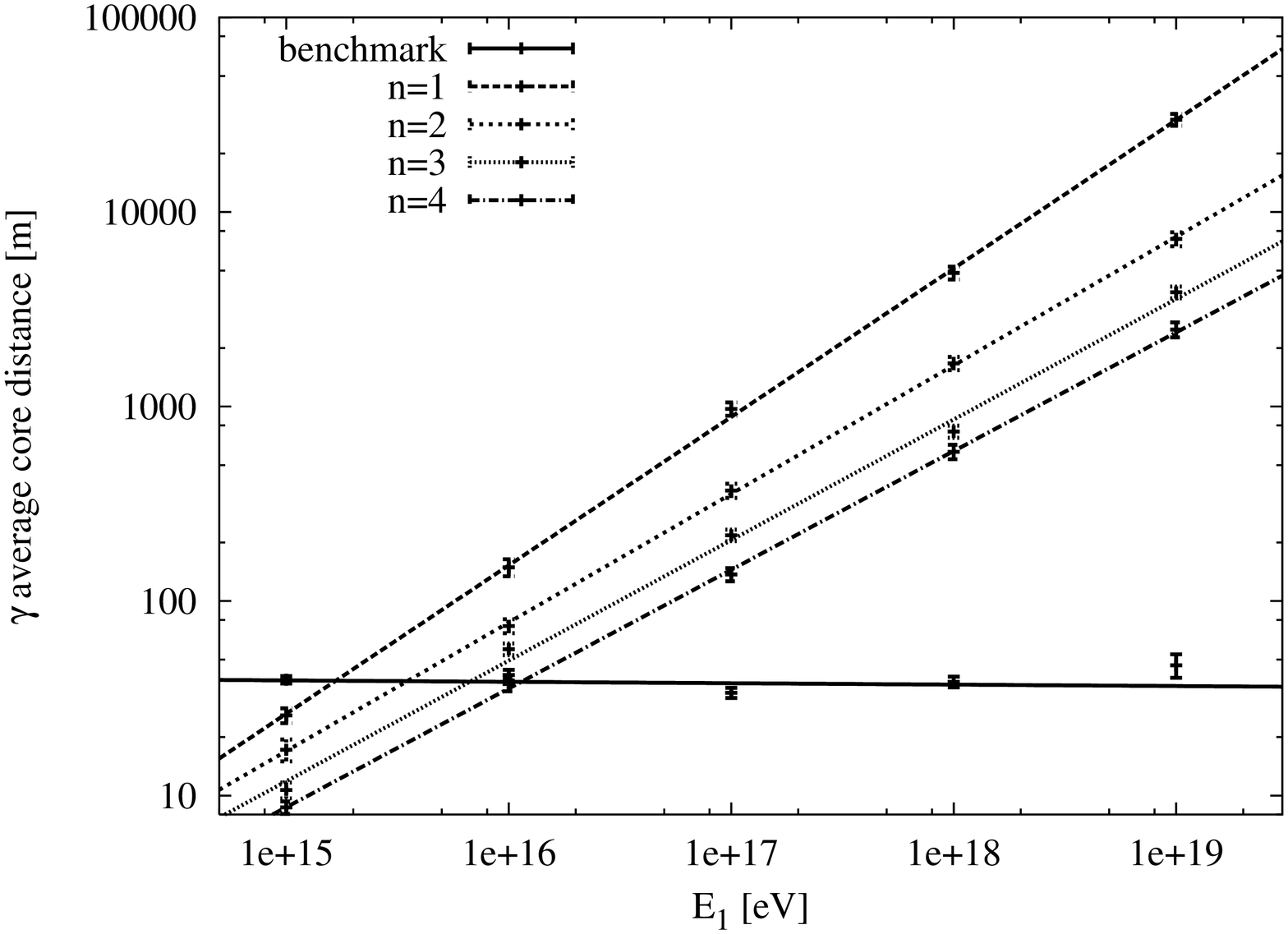}}} \par}
\caption{As in Fig. \ref{terza}, but this time the first interaction occurs at \protect\( 15000\, \textrm{m}\protect \)
(124 $g/cm^2$).}
\label{quarta}
\end{figure}

As we have explained in \cite{CCT}, a shortcoming of CORSIKA 
is that the program is not able to provide a neutrino primary, and for 
this reason we have used as benchmarks simulations of proton-proton collisions 
with the same energy and at the same altitude of the signal black hole events. 
For this reason, the multiplicities 
of the benchmark events provide an upper bound for the multiplicities of a 
real event, since hadronic events 
have clearly larger multiplicities compared, for instance, to neutrino events. 
As we increase the altitude (Fig.~\ref{seconda}) 
benchmark events and black hole events have similar slopes and intercepts, 
with minor differences.      
Independently of the altitude, black hole events are characterized by wider 
lateral distributions (Figs.~\ref{terza} and \ref{quarta})
compared to the benchmark result. This is due to the fireball structure 
of the black hole event, with its original 
s-wave emission. Our results, of course, rely on a semiclassical
picture of the particle emission by the black hole, given by the Hawking formula, 
and we have corrected for gravity loss in the bulk as well. 

We summarize by saying that an 
analysis of lateral distributions and total multiplicities 
as a function of energy show a simple
linear behaviour (in a log/log scale plot) and measurement of these 
observables and study of their correlation will be useful criteria
to discern among various models for the underlying interactions.

\section{PERSPECTIVES}

With AUGER being now operational \cite{auger}, we do expect that a systematic 
analysis of extensive air showers will provide 
some evidence, in the near future, 
for excluding or discovering a low gravity scale, based on experimental data. 
This will probably render the field more mature from the phenomenological 
side. It is also likely that most of these studies will not be conclusive, 
in this respect, but we will be able to set an improved 
lower bound on the value of $M_*$, the low gravity scale, which, unfortunately, 
is not a prediction but just a parameter of the theory. 
\centerline{\bf Note Added}
After completing this investigation we have been informed by D. Heck that 
a new version of CORSIKA has been released which allows to treat neutrino 
as primaries in the simulations. As we have discussed above, the use 
of neutrino as a benchmark simulation should only render more remarked
the differences between black hole events and benchmark events in the 
characteristics of the air showers that have been studied here.

\end{document}